\documentclass[prd,showpacs,showkeys,floatfix,twocolumn,amsmath,amssymb,floatfix]{revtex4}
\usepackage{graphicx,color,dcolumn,booktabs,bm}
\usepackage{longtable,lscape}
\usepackage{amssymb}
\usepackage{indentfirst}
\usepackage{feynmf}   
\usepackage{epstopdf}   
\usepackage{slashed}  
\usepackage{cases}
\usepackage{color}
\usepackage{multirow}
\usepackage[colorlinks,citecolor=blue,anchorcolor=red,menucolor=red,linkcolor=red,filecolor=red,runcolor=red,urlcolor=blue,frenchlinks=red]{hyperref}
\usepackage{dcolumn}
\usepackage{bm}
\usepackage{enumerate}
\allowdisplaybreaks[4]

\def\FF(s){\left[(\alpha+\beta)m_c^2-\alpha\beta s\right]}
\def\HH(s){\left[m_c^2-\alpha(1-\alpha) s\right]}

\begin{document}

\title{A suggested search for doubly charmed baryons of $J^P=3/2^+$ \\ via their electromagnetic transitions}

\author{Er-Liang Cui$^1$}
\author{Hua-Xing Chen$^1$}
\email{hxchen@buaa.edu.cn}
\author{Wei Chen$^2$}
\email{wec053@mail.usask.ca}
\author{Xiang Liu$^{3,4}$}
\email{xiangliu@lzu.edu.cn}
\author{Shi-Lin Zhu$^{5,6,7}$}
\email{zhusl@pku.edu.cn}
\affiliation{
$^1$School of Physics and Beijing Key Laboratory of Advanced Nuclear Materials and Physics, Beihang University, Beijing 100191, China \\
$^2$School of Physics, Sun Yat-Sen University, Guangzhou 510275, China \\
$^3$School of Physical Science and Technology, Lanzhou University, Lanzhou 730000, China\\
$^4$Research Center for Hadron and CSR Physics, Lanzhou University and Institute of Modern Physics of CAS, Lanzhou 730000, China\\
$^5$School of Physics and State Key Laboratory of Nuclear Physics and Technology, Peking University, Beijing 100871, China \\
$^6$Collaborative Innovation Center of Quantum Matter, Beijing 100871, China \\
$^7$Center of High Energy Physics, Peking University, Beijing 100871, China}

\begin{abstract}
We use the method of light-cone sum rules to study the electromagnetic transition of the $\Xi^{*++}_{cc}$ into $\Xi^{++}_{cc}\gamma$, whose decay width is estimated to be $13.7~{^{+17.7}_{-~7.9}}$ keV. This value is large enough for the $\Xi^{*++}_{cc}$ to be observed in the $\Xi^{++}_{cc}\gamma$ channel, and we propose to continually search for it in future LHCb and BelleII experiments.
\end{abstract}
\pagenumbering{arabic}
\maketitle

\section{Introduction}
\label{sec:intro}

The doubly heavy baryons provide an ideal platform to study the heavy quark symmetry, and have been investigated in various experimental and theoretical studies during the past three decades~\cite{pdg}. Experimentally, in 2002 the SELEX Collaboration reported the evidence of the doubly charmed baryon $\Xi_{cc}^+(3519)$ in the $\Xi_{cc}^+\to \Lambda_c^+ K^- \pi^+$ process and determined its mass to be $3518.9\pm0.9$ MeV~\cite{Mattson:2002vu}. However, all the other experiments did not confirm this~\cite{experiment}. Theoretically, lots of methods and models have been applied to study the doubly charmed baryons, such as the bag model~\cite{bagmodel}, various quark models~\cite{quarkmodel}, QCD sum rules~\cite{lattice}, and lattice QCD~\cite{sumrule}, etc.~\cite{others}. We refer to reviews~\cite{review} for more relevant discussions.

Recently, the doubly heavy baryon $\Xi^{++}_{cc}(3621)$ was discovered by the LHCb Collaboration in the $\Lambda^+_c K^- \pi^+ \pi^+$ mass spectrum~\cite{Aaij:2017ueg}, which channel was previously suggested by F.~S.~Yu {\it et al.} in Ref.~\cite{Yu:2017zst}. The LHCb experiment measured its mass to be
\begin{eqnarray}
M_{\Xi^{++}_{cc}} = 3621.40 \pm 0.72 \pm 0.27 \pm 0.14~{\rm MeV} \, ,
\end{eqnarray}
which value is significantly larger than the mass of the $\Xi_{cc}^+(3519)$ determined by SELEX~\cite{Mattson:2002vu}. Because the $\Xi_{cc}^{++}$ and $\Xi_{cc}^{+}$ are isospin partners whose mass difference should be only a few MeV,
the LHCb experiment~\cite{Aaij:2017ueg} did not confirm the SELEX experiment~\cite{Mattson:2002vu} neither, but discovered a new state.

The discovery of the $\Xi^{++}_{cc}(3621)$ quickly attracted much attention from the hadron physics community, and many theoretical methods were applied to study it~\cite{other}.
Especially, its weak decay properties were studied in Refs.~\cite{weak}, its magnetic moments were studied in Refs.~\cite{magnetic}, and its relevant molecular states were investigated in Refs.~\cite{molecule}.

Besides the mass of the $\Xi^{++}_{cc}(3621)$, the LHCb experiment preferentially retains longer-lived $\Xi^{++}_{cc}$ candidates and favors the $J^P = 1/2^+$ assignment~\cite{Aaij:2017ueg}.
Hence, it is natural to continually search for the doubly charmed baryon $\Xi^{*++}_{cc}$ of $J^P = 3/2^+$.
One of its possible decay channels is the radiative decay $\Xi^{*++}_{cc} \rightarrow \Xi^{++}_{cc}\gamma$.
This has been investigated and the decay width of $\Xi^{*++}_{cc} \rightarrow \Xi^{++}_{cc}\gamma$ was evaluated to be a few or tens of keV when using various phenomenological models~\cite{Hackman:1977am,Bernotas:2013eia,Xiao:2017udy,Branz:2010pq,Lu:2017meb,Li:2017pxa}.
Very probably, these values are much larger than the weak decay width of the $\Xi^{*++}_{cc}$, making it possible and promising to search for the $\Xi^{*++}_{cc}$ in the $\Xi^{++}_{cc}\gamma$ channel.

To further verify the above results, in this paper we use the method of light-cone sum rules to study the electromagnetic transition of the $\Xi^{*++}_{cc}$ into $\Xi^{++}_{cc}\gamma$, based on our previous study on the mass spectrum of doubly charmed baryons using QCD sum rules~\cite{Chen:2017sbg}. This paper is organized as follows. In Sec.~\ref{sec:sumrule} we shall use the light-cone sum rules to study the electromagnetic transition of the $\Xi^{*++}_{cc}$ into $\Xi^{++}_{cc}\gamma$. The numerical analyses will be done in Sec.~\ref{sec:numerical}, and the results will be summarized and discussed in Sec.~\ref{sec:summary}.

\section{Light-cone sum rules}
\label{sec:sumrule}


In this section we use the method of light-cone sum rules to study the electromagnetic transition of the $\Xi^{*++}_{cc}$ into $\Xi^{++}_{cc}\gamma$, which method has been widely used to study decay properties of hadrons~\cite{lightcone,Ball:2002ps,ball}.
We have also systematically studied mass spectra and decay properties of singly charmed baryons using QCD sum rules and light-cone sum rules in the framework of heavy quark effective theory~\cite{hqetsumrule}.

To use this method we first investigate the following three-point correlation function:
\begin{eqnarray}
&& \Pi_\alpha(p, \, k , \, q , \, \epsilon)
\label{eq:correlation}
\\ \nonumber && ~~~~~~~~ = \int d^4 x e^{-i k \cdot x} \langle 0 | J_{\Xi^{*++}_{cc},\alpha}(0) \bar J_{\Xi^{++}_{cc}}(x) | \gamma(q,\epsilon) \rangle \, ,
\end{eqnarray}
where $p (= k+q)$, $k$, $q$ are the momenta of the $\Xi^{*++}_{cc}$, $\Xi^{++}_{cc}$, and $\gamma$, respectively; $\epsilon$ is the polarization vector of the $\gamma$. 
Note that we have interchanged the initial and final states to be $\Xi^{++}_{cc} \gamma \rightarrow \Xi^{*++}_{cc}$ in the above equation.
The currents $J_{\Xi^{++}_{cc}}$ and $J_{\Xi^{*++}_{cc},\alpha}$ have been given in Ref.~\cite{Chen:2017sbg}:
\begin{eqnarray}
J_{\Xi^{++}_{cc}}(x) &=& \cos \theta_1 \times \eta_1(x) + \sin \theta_1 \times \eta_{1}^\prime(x)
\label{def:eta1mix}
\\ \nonumber &=& \cos \theta_1 \times \epsilon^{abc} \left(c_a^T(x) C \gamma_\mu c_b(x)\right) \gamma^\mu \gamma_5 q_c(x)
\\ \nonumber &+& \sin \theta_1 \times \epsilon^{abc} \left(q_a^T(x) C \gamma_5 c_b(x)\right) c_c(x)
\\ \nonumber &=& t_1 \times \epsilon^{abc} \left(c_a^T(x) C \gamma_\mu c_b(x)\right) \gamma^\mu \gamma_5 q_c(x)
\\ \nonumber &+& t_2 \times \epsilon^{abc} \left(c_a^T(x) C \sigma_{\mu\nu} c_b(x)\right) \sigma^{\mu\nu} \gamma_5 q_c(x)  \, ,
\\
J_{\Xi^{*++}_{cc},\alpha}(x) &=& \cos \theta_2 \times \eta_{3\alpha}(x) + \sin \theta_2 \times \eta_{3\alpha}^\prime(x)
\label{def:eta2mix}
\\ \nonumber &=& \cos \theta_2 \times \Gamma_{\alpha\mu} \epsilon^{abc} \left(c_a^T(x) C \gamma^{\mu} c_b(x)\right) q_c(x)
\\ \nonumber &+& \sin \theta_2 \times \Gamma_{\alpha\mu} \epsilon^{abc} \left(q_a^T(x) C \gamma^{\mu} c_b(x)\right) c_c(x) \, ,
\end{eqnarray}
where
\begin{eqnarray}
t_1 &=& \cos \theta_1 - {\sin \theta_1 \over 4} \, ,
\\ \nonumber t_2 &=& - { \sin \theta_1 \over 8} \, ,
\end{eqnarray}
and $\Gamma_{\alpha\mu}$ is the projection operator
\begin{eqnarray}
\Gamma_{\mu\nu} = g_{\mu\nu} - {1\over4}\gamma_\mu\gamma_\nu \, .
\end{eqnarray}
The above two currents couple to the $\Xi^{++}_{cc}$ and $\Xi^{*++}_{cc}$ through:
\begin{eqnarray}
\langle 0 | J_{\Xi^{++}_{cc}} | \Xi^{++}_{cc} \rangle &=& f_{\Xi^{++}_{cc}} u_{{\Xi}^{++}_{cc}}(p) \, ,
\label{eq:coupling1}
\\ \langle 0 | J_{\Xi^{*++}_{cc},\alpha} | \Xi^{*++}_{cc} \rangle &=& f_{\Xi^{*++}_{cc}} u_{{\Xi}^{*++}_{cc},\alpha}(p) \, .
\label{eq:coupling2}
\end{eqnarray}

At the hadronic level, we write the amplitude of the $\Xi^{*++}_{cc} \rightarrow \Xi^{++}_{cc} \gamma$ as
\begin{eqnarray}
\mathcal{M}_{\Xi^{*++}_{cc} \rightarrow \Xi^{++}_{cc} \gamma}
\label{eq:amplitude} = e~g~\epsilon^{\mu\nu\rho\sigma}~\overline{u}_{\Xi^{++}_{cc}}~{u_{{\Xi}^{*++}_{cc},\rho}}~p_\mu q_\nu \epsilon^*_\sigma \, ,
\end{eqnarray}
where $g \equiv g_{\Xi^{*++}_{cc} \rightarrow \Xi^{++}_{cc} \gamma}$ is the coupling constant and $e$ is the charge of the proton. Inserting Eqs.~(\ref{eq:coupling1}-\ref{eq:amplitude}) into Eq.~(\ref{eq:correlation}), we find $\Pi_\alpha(p, \, k , \, q , \, \epsilon)$ has the following pole terms:
\begin{eqnarray}
&& \Pi_\alpha(p, \, k , \, q , \, \epsilon)
\label{eq:hadron}
\\ \nonumber &=& e~g~\epsilon^{\mu\nu\rho\sigma}p_\mu q_\nu \epsilon_\sigma \times {f_{\Xi^{*++}_{cc}} f_{\Xi^{++}_{cc}} \over (p^2 - M^2_{{\Xi}^{*++}_{cc}}) (k^2 - M^2_{{\Xi}^{++}_{cc}})}
\\ \nonumber && ~~~~~ \times \left( g_{\alpha\rho} - {1\over3} \gamma_\alpha \gamma_\rho - {p_\alpha \gamma_\rho - p_\rho \gamma_\alpha \over 3M_{{\Xi}^{*++}_{cc}}} - {2p_\alpha p_\rho \over 3M_{{\Xi}^{*++}_{cc}}^2} \right)
\\ \nonumber && ~~~~~ \times \left( p\!\!\!\slash + M_{{\Xi}^{*++}_{cc}} \right) \left( k\!\!\!\slash + M_{{\Xi}^{++}_{cc}} \right)
\\ \nonumber &\approx& {2e\over3}~g~\epsilon^{\alpha\nu\rho\sigma}p_\nu q_\rho \epsilon_\sigma \times {f_{\Xi^{*++}_{cc}} f_{\Xi^{++}_{cc}} \over (p^2 - \widetilde M^2)^2} \times \left( p^2 + \widetilde M^2 \right)
\\ \nonumber && ~~~~~ + \cdots \, ,
\end{eqnarray}
where its leading component is obtained after assuming that $p \approx k \gg q$, $M_{{\Xi}^{++}_{cc}} \approx M_{{\Xi}^{*++}_{cc}}$, and $\widetilde M \equiv (M_{{\Xi}^{++}_{cc}} + M_{{\Xi}^{*++}_{cc}})/2$.
Note that we have kept only the double-pole term but omitted the single-pole terms, which gives some but not large uncertainties.
Here we have used the following formula for the baryon fields of spin 1/2 and 3/2:
\begin{eqnarray}
&& \sum_{spin} u(p) \bar u(p) = p\!\!\!\slash + m \, ,
\\ && \sum_{spin} u_{\mu}(p) \bar u_{\nu}(p)
\\ \nonumber &=& \left( g_{\mu\nu} - {1\over3} \gamma_\mu \gamma_{\nu} - {p_\mu\gamma_{\nu} - p_{\nu}\gamma_{\mu} \over 3m} - {2p_{\mu}p_{\nu} \over 3m^2} \right) \left( p\!\!\!\slash + m \right) \, .
\end{eqnarray}

At the quark and gluon level, we calculate $\Pi_\alpha(p, \, k , \, q , \, \epsilon)$ using the method of operator product expansion (OPE)~\cite{Mertig:1990an}. The result is quite messy, so we show it in Eq.~(\ref{eq:sumrule1}) in Appendix~\ref{sec:result}.
It can be naturally separated into two parts, i.e., the photon can be omitted either from the light quark or from the two charm quarks.
There are many light-cone photon distribution amplitudes contained in that equation, whose definitions can be found in Ref.~\cite{Ball:2002ps}.
For completeness, they are also listed in Appendix~\ref{sec:wavefunction}.

Finally, we use the approximation $p \approx k \gg q$ once more to write the two coefficients $e^{-i k x} e^{- i (1-u) qx}$ and $e^{-i k x} e^{-i (\alpha_2 + w \alpha_3) qx}$ as $e^{-i p x}$. After performing the Borel transformation at both the hadron and quark-gluon levels, we obtain the sum rules as shown in Eq.~(\ref{eq:sumrule2}) in Appendix~\ref{sec:result}, where $s_0$ and $M_B$ are the threshold value and Borel mass, respectively.

Similarly, we calculate the sum rules related to electromagnetic transitions of some other doubly charmed and bottom baryons, i.e., $\Xi^{*+}_{cc} \rightarrow \Xi^{+}_{cc}\gamma$, $\Omega^{*+}_{cc} \rightarrow \Omega^{+}_{cc}\gamma$, $\Xi^{*0}_{bb} \rightarrow \Xi^{0}_{bb}\gamma$, $\Xi^{*-}_{bb} \rightarrow \Xi^{-}_{bb}\gamma$, and $\Omega^{*-}_{bb} \rightarrow \Omega^{-}_{bb}\gamma$.
We shall use them to perform numerical analyses in the next section.

\section{Numerical Analyses}
\label{sec:numerical}

To perform numerical analyses, we use the following values at the energy scale $\mu = 1$ GeV:
\begin{enumerate}

\item The parameters in the distribution amplitudes take the following values~\cite{Ball:2002ps,Balitsky:1997wi}:
\begin{eqnarray}
\nonumber && f_{3\gamma} = -(4 \pm 2)\cdot10^{-3}~{\rm GeV}^2 \, ,
\\ \nonumber && \omega^V_\gamma = 3.8 \pm 1.8 \, ,
\\ && \omega^A_\gamma = -2.1 \pm 1.0 \, ,
\\ \nonumber && \chi = (3.15\pm0.10)~{\rm GeV}^{-2} \, ,
\\ \nonumber && \kappa = 0.2 \, , \zeta_1 = 0.4 \, , \zeta_2 = 0.3 \, ,
\\ \nonumber && \kappa^+ = \zeta^+_1 = \zeta^+_2 = 0 \, .
\end{eqnarray}
Note that we use the above values for both the vector mesons $\rho$ and $\phi$, which can give some but not large uncertainties.

\item The quark and gluon condensates take the following values~\cite{pdg,value2}:
\begin{eqnarray}
\nonumber && \langle \bar qq \rangle = - (0.24 \pm 0.01)^3 \mbox{ GeV}^3 \, ,
\\ \nonumber && \langle \bar ss \rangle = 0.8 \times \langle\bar qq \rangle \, ,
\\ \nonumber &&\langle g_s^2GG\rangle =(0.48\pm 0.14) \mbox{ GeV}^4\, ,
\\ \label{condensates} && \langle g_s \bar q \sigma G q \rangle = M_0^2 \times \langle \bar qq \rangle\, ,
\\ \nonumber && \langle g_s \bar s \sigma G s \rangle = M_0^2 \times \langle \bar ss \rangle\, ,
\\ \nonumber && m_s = 96^{+8}_{-4} \mbox{ MeV} \, ,
\\ \nonumber && m_c = 1.23 \pm 0.09 \mbox{ GeV} \, ,
\\ \nonumber && m_b = 4.18 ^{+0.04}_{-0.03} \mbox{ GeV} \, .
\end{eqnarray}

\item There are many parameters related to the doubly charmed interpolating currents $J_{\Xi^{++}_{cc}}$ and $J_{\Xi^{*++}_{cc},\alpha}$. We list them in Table~\ref{tab:currents} together with those related to $J_{\Omega^{+}_{cc}}$ and $J_{\Omega^{*+}_{cc},\alpha}$. The parameters related to $J_{\Xi^{+}_{cc}}$ and $J_{\Xi^{*+}_{cc},\alpha}$ are the same as those related to $J_{\Xi^{++}_{cc}}$ and $J_{\Xi^{*++}_{cc},\alpha}$. All these values are taken from Ref.~\cite{Chen:2017sbg}, where the mass spectrum of doubly charmed baryons is investigated by using the method of QCD sum rules.

    We also use the same QCD sum rule method to evaluate the parameters related to their bottom partners, i.e., the $S$-wave doubly charmed baryons $\Xi^{(*)}_{bb}$ and $\Omega^{(*)}_{bb}$. The extracted masses are listed in last column of Table~\ref{tab:currents}, which are consistent with Ref.~\cite{Ebert:2002ig}, where the mass spectrum of doubly heavy baryons is investigated by using the relativistic quark model. Note that although we list the mass values evaluated using QCD sum rules, we do not use them for numerical analyses in the present study.

\item The mass of the $\Xi^{++}_{cc}$ has been measured by the LHCb Collaboration to be $M_{\Xi^{++}_{cc}} = 3621.40 \pm 0.72 \pm 0.27 \pm 0.14~{\rm MeV}$~\cite{Aaij:2017ueg}. For the masses of the $\Xi^{*++}_{cc}$ and other doubly charmed baryons, we take the values from Ref.~\cite{Ebert:2002ig}, where the mass spectrum of doubly heavy baryons is investigated by using the relativistic quark model. We list all these values in the second column of Table~\ref{tab:currents}, which will be used for numerical analyses in the following.

\end{enumerate}

\renewcommand{\arraystretch}{1.5}
\begin{table*}[hbtp]
\begin{center}
\caption{Parameters related to doubly charmed baryons and their relevant interpolating currents, taken from Refs.~\cite{Aaij:2017ueg,Chen:2017sbg,Ebert:2002ig}. The parameters related to $J_{\Xi^{+}_{cc}}$ and $J_{\Xi^{*+}_{cc},\alpha}$ are the same as those related to $J_{\Xi^{++}_{cc}}$ and $J_{\Xi^{*++}_{cc},\alpha}$. Note that we use the mass values listed in the second column for numerical analyses. The mass values listed in the last (eighth) column are evaluated using the QCD sum rule method, which we do not use for numerical analyses in the present study.}
\begin{tabular}{cc|cccccc}
\toprule[1pt]\toprule[1pt]
~~\mbox{Baryon}~~ & ~~\mbox{Mass~(GeV)}~~ & ~~\mbox{Currents}~~ & ~~\mbox{Mixing Angle}~~ & ~~$s_0$~(GeV$^2$)~~ & ~~$M_B^2$~(GeV$^2$)~~ & ~~$f$~(GeV$^3$)~~ & ~~\mbox{Mass~(GeV)}~\cite{Chen:2017sbg}~~
\\ \midrule[1pt]
$\Xi^{++}_{cc}$ & $3621.40$~\cite{Aaij:2017ueg} & $J_{\Xi^{++}_{cc}}$ & $\theta_1 = - 11\pm5^\circ$ & $25\pm3$ & $3.5\pm0.3$  &  $0.15^{+0.02}_{-0.03}$  &  $3.58^{+0.15}_{-0.16}$
\\
$\Xi^{*++}_{cc}$ & $3727$~\cite{Ebert:2002ig} & $J_{\Xi^{*++}_{cc}}$ & $\theta_2 = 6\pm3^\circ$ & $25\pm3$ & $3.5\pm0.3$  &  $0.061^{+0.006}_{-0.009}$  &  $3.58^{+0.14}_{-0.10}$
\\
$\Omega^{+}_{cc}$ & $3778$~\cite{Ebert:2002ig} & $J_{\Omega^{+}_{cc}}$ & $\theta_1 = - 11\pm5^\circ$ & $25\pm3$ & $3.5\pm0.3$  &  $0.17^{+0.02}_{-0.03}$  &  $3.70^{+0.13}_{-0.15}$
\\
$\Omega^{*+}_{cc}$ & $3872$~\cite{Ebert:2002ig} & $J_{\Omega^{*+}_{cc}}$ & $\theta_2 = 6\pm3^\circ$ & $25\pm3$ & $3.5\pm0.3$  &  $0.074^{+0.008}_{-0.011}$  &  $3.69^{+0.11}_{-0.15}$
\\
$\Xi^{0}_{bb}$ & $10202$~\cite{Ebert:2002ig} & $J_{\Xi^{++}_{bb}}$ & $\theta_1 = - 11\pm5^\circ$ & $130\pm10$ & $28\pm3$  &  $0.67^{+0.20}_{-0.20}$  & $10.37^{+0.31}_{-0.34}$
\\
$\Xi^{*0}_{bb}$ & $10237$~\cite{Ebert:2002ig} & $J_{\Xi^{*++}_{bb}}$ & $\theta_2 = 6\pm3^\circ$ & $135\pm10$ & $34\pm3$  &  $0.29^{+0.08}_{-0.07}$  & $10.47^{+0.32}_{-0.32}$
\\
$\Omega^{-}_{bb}$ & $10359$~\cite{Ebert:2002ig} & $J_{\Omega^{+}_{bb}}$ & $\theta_1 = - 11\pm5^\circ$ & $130\pm10$ & $28\pm3$  &  $0.82^{+0.22}_{-0.22}$  & $10.45^{+0.29}_{-0.32}$
\\
$\Omega^{*-}_{bb}$ & $10389$~\cite{Ebert:2002ig} & $J_{\Omega^{*+}_{bb}}$ & $\theta_2 = 6\pm3^\circ$ & $135\pm10$ & $34\pm3$  &  $0.38^{+0.09}_{-0.09}$  & $10.60^{+0.29}_{-0.31}$
\\ \bottomrule[1pt]\bottomrule[1pt]
\end{tabular}
\label{tab:currents}
\end{center}
\end{table*}

Inserting the above values into the sum rule (\ref{eq:sumrule2}), we evaluate the coupling constant $g_{\Xi^{*++}_{cc} \rightarrow \Xi^{++}_{cc} \gamma}$ to be
\begin{eqnarray}
&&g_{\Xi^{*++}_{cc} \rightarrow \Xi^{++}_{cc} \gamma} = 0.30~{^{+0.16}_{-0.11}}~{\rm GeV}^{-2}
\\ \nonumber &&= 0.304~{^{+0.024}_{-0.026}}~{^{+0.007}_{-0.014}}~{^{+0.033}_{-0.028}}~{^{+0.093}_{-0.045}}~{^{+0.118}_{-0.087}}~{\rm GeV}^{-2} \, ,
\end{eqnarray}
where the uncertainties are due to the mixing angles $\theta_{1/2}$, the threshold value $s_0$, the Borel mass $M_B$, the decay constants $f_{\Xi^{++}_{cc}}$ and $f_{\Xi^{*++}_{cc}}$, and various condensates and parameters in the OPE series and distribution amplitudes, respectively.

\begin{figure}[htb]
\begin{center}
\scalebox{0.6}{\includegraphics{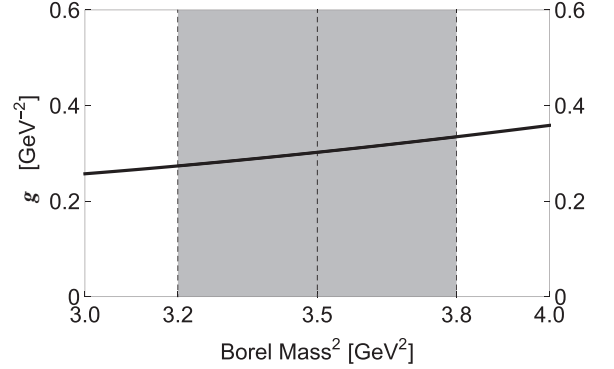}}
\end{center}
\caption{The coupling constant $g_{\Xi^{*++}_{cc} \rightarrow \Xi^{++}_{cc} \gamma}$ as a function of the Borel mass $M_B$.
\label{fig:coupling}}
\end{figure}

For completeness, we show the coupling constant $g_{\Xi^{*++}_{cc} \rightarrow \Xi^{++}_{cc} \gamma}$ as a function of the Borel mass in Fig.~\ref{fig:coupling}, and find that the curve is quite stable inside the Borel window $3.2$ GeV$^2< M_B^2 < 3.8$ GeV$^2$.

Finally, we use the following decay formula
\begin{equation}
\Gamma_{\Xi^{*++}_{cc} \rightarrow \Xi^{++}_{cc} \gamma} = {|~{\vec q}~| \over 8\pi M^2_{\Xi^{*++}_{cc}}} \times | {1\over4} \sum_{\rm spin} \mathcal{M}_{\Xi^{*++}_{cc} \rightarrow \Xi^{++}_{cc} \gamma} |^2 \, ,
\end{equation}
to obtain
\begin{eqnarray}
\Gamma_{\Xi^{*++}_{cc} \rightarrow \Xi^{++}_{cc} \gamma} &=& 13.7~{^{+17.7}_{-~7.9}} {\rm~keV} \, ,
\label{eq:gamma}
\end{eqnarray}
where $|{\vec q}|$ is the momentum of the photon in the rest frame of the $\Xi^{*++}_{cc}$.
Note that the uncertainties of the masses of the $\Xi^{++}_{cc}$ and $\Xi^{*++}_{cc}$ have not been taken into account. Neither do some of the parameters contained in the distribution amplitudes, such as $\kappa$ and $\kappa^+$, etc.
Moreover, we have made some approximations when deriving Eq.~(\ref{eq:hadron}). Therefore, the total uncertainties can be larger and the final result can be three times larger or smaller than those we have obtained, i.e.,
$13.7~^{+200\%}_{-~67\%}$ keV, while we shall still use Eq.~(\ref{eq:gamma}) as our prediction.

Using the same approach, we calculate the following coupling constants
\begin{eqnarray}
\nonumber g_{\Xi^{*+}_{cc} \rightarrow \Xi^{+}_{cc} \gamma} &=& 0.23~{^{+0.13}_{-0.09}} {\rm~GeV}^{-2} \, ,
\\ \nonumber g_{\Omega^{*+}_{cc} \rightarrow \Omega^{+}_{cc} \gamma} &=& 0.22~{^{+0.11}_{-0.08}} {\rm~GeV}^{-2} \, ,
\\ g_{\Xi^{*0}_{bb} \rightarrow \Xi^{0}_{bb}\gamma} &=& 0.050~{^{+0.026}_{-0.018}} {\rm~GeV}^{-2} \, ,
\\ \nonumber g_{\Xi^{*-}_{bb} \rightarrow \Xi^{-}_{bb}\gamma} &=& 0.081~{^{+0.031}_{-0.020}} {\rm~GeV}^{-2} \, ,
\\ \nonumber g_{\Omega^{*-}_{bb} \rightarrow \Omega^{-}_{bb}\gamma} &=& 0.053~{^{+0.015}_{-0.013}} {\rm~GeV}^{-2} \, ,
\end{eqnarray}
and estimate their relevant electromagnetic transitions to be
\begin{eqnarray}
\nonumber \Gamma_{\Xi^{*+}_{cc} \rightarrow \Xi^{+}_{cc} \gamma} &=& 8.1~{^{+11.1}_{-~4.9}} {\rm~keV} \, ,
\\ \nonumber \Gamma_{\Omega^{*+}_{cc} \rightarrow \Omega^{+}_{cc} \gamma} &=& 5.4~{^{+6.9}_{-3.1}} {\rm~keV} \, ,
\\ \Gamma_{\Xi^{*0}_{bb} \rightarrow \Xi^{0}_{bb}\gamma} &=& 0.11~{^{+0.13}_{-0.07}} {\rm~keV} \, ,
\\ \nonumber \Gamma_{\Xi^{*-}_{bb} \rightarrow \Xi^{-}_{bb}\gamma} &=& 0.28~^{+0.24}_{-0.13} {\rm~keV} \, ,
\\ \nonumber \Gamma_{\Omega^{*-}_{bb} \rightarrow \Omega^{-}_{bb}\gamma} &=& 0.08~^{+0.05}_{-0.04} {\rm~keV} \, .
\end{eqnarray}
We list all the above decay widths in Table~\ref{tab:result}.

\renewcommand{\arraystretch}{1.5}
\begin{table*}[hbtp]
\begin{center}
\caption{Electromagnetic transitions of doubly charmed baryons, in units of keV. For comparison, we also list the results obtained using
the bag model~\cite{Hackman:1977am,Bernotas:2013eia},
the nonrelativistic constituent quark model~\cite{Xiao:2017udy},
the relativistic constituent three-quark model including hyperfine mixing effects~\cite{Branz:2010pq},
the relativistic constituent quark model within the diquark picture~\cite{Lu:2017meb},
and the chiral perturbation theory~\cite{Li:2017pxa}.}
\begin{tabular}{cccccccc}
\toprule[1pt]\toprule[1pt]
~~\mbox{Process}~~ & ~~\mbox{Our results}~~ & ~~~Ref.~\cite{Hackman:1977am}~~~ & ~~~Ref.~\cite{Bernotas:2013eia}~~~ & ~~~Ref.~\cite{Xiao:2017udy}~~~ & ~~~Ref.~\cite{Branz:2010pq}~~~ & ~~~Ref.~\cite{Lu:2017meb}~~~ & ~~~Ref.~\cite{Li:2017pxa}~~~
\\ \midrule[1pt]
~~$\Xi^{*++}_{cc} \rightarrow \Xi^{++}_{cc} \gamma$~~ & $13.7~{^{+17.7}_{-~7.9}}$ &  $4.35$ & $1.43$ & $16.7$ & $23.46\pm3.33$ & $7.21$   & $22.0$
\\
~~$\Xi^{*+}_{cc} \rightarrow \Xi^{+}_{cc} \gamma$~~ & $8.1~{^{+11.1}_{-~4.9}}$ & $3.96$ & $2.08$ & $14.6$ & $28.79\pm2.51$ & $3.90$ & $9.57$
\\
~~$\Omega^{*+}_{cc} \rightarrow \Omega^{+}_{cc} \gamma$~~   & $5.4~{^{+6.9}_{-3.1}}$ &  $1.35$  & $0.95$  & $6.93$  & $2.11\pm0.11$  & $0.82$ &  $9.45$
\\ \hline
~~$\Xi^{*0}_{bb} \rightarrow \Xi^{0}_{bb} \gamma$~~   & $0.11~{^{+0.13}_{-0.07}}$ &   -- & --  & $1.19$  & $0.31\pm0.06$  & $0.98$ &  --
\\
~~$\Xi^{*-}_{bb} \rightarrow \Xi^{-}_{bb} \gamma$~~   & $0.28~^{+0.24}_{-0.13}$ &  --  & --  &  $0.24$ & $0.0587\pm0.0142$  & $0.21$ &  --
\\
~~$\Omega^{*-}_{bb} \rightarrow \Omega^{-}_{bb} \gamma$~~   & $0.08~^{+0.05}_{-0.04}$ &  --  & --  &  $0.08$ & $0.0226\pm0.0045$  & $0.04$ &  --
\\ \bottomrule[1pt]\bottomrule[1pt]
\end{tabular}
\label{tab:result}
\end{center}
\end{table*}


\section{Summary and Discussions}
\label{sec:summary}

In this paper we have used the method of light-cone sum rules to study the electromagnetic transition of the $\Xi^{*++}_{cc}$ into $\Xi^{++}_{cc}\gamma$, whose decay width is estimated to be
\begin{eqnarray}
\Gamma_{\Xi^{*++}_{cc} \rightarrow \Xi^{++}_{cc} \gamma} &=& 13.7~{^{+17.7}_{-~7.9}} {\rm~keV} \, .
\end{eqnarray}
We have also investigated electromagnetic transitions of some other doubly charmed and bottom baryons, including $\Xi^{*+}_{cc} \rightarrow \Xi^{+}_{cc}\gamma$, $\Omega^{*+}_{cc} \rightarrow \Omega^{+}_{cc}\gamma$, $\Xi^{*0}_{bb} \rightarrow \Xi^{0}_{bb}\gamma$, $\Xi^{*-}_{bb} \rightarrow \Xi^{-}_{bb}\gamma$, and $\Omega^{*-}_{bb} \rightarrow \Omega^{-}_{bb}\gamma$. To study the latter three processes, we have used the method of QCD sum rules to investigate the mass spectrum of doubly bottom baryons. The extracted masses are listed in last column of Table~\ref{tab:currents}, consistent with Ref.~\cite{Ebert:2002ig} where the mass spectrum of doubly heavy baryons is investigated by using the relativistic quark model.

We summarized the above results in Table~\ref{tab:result} together with those evaluated in Refs.~\cite{Hackman:1977am,Bernotas:2013eia,Xiao:2017udy,Branz:2010pq,Lu:2017meb,Li:2017pxa}.
Our results are (slightly) larger than those obtained using the bag model~\cite{Hackman:1977am,Bernotas:2013eia}, but quite comparable with those obtained using the nonrelativistic constituent quark model~\cite{Xiao:2017udy},
the relativistic three-quark model including hyperfine mixing effects~\cite{Branz:2010pq}, the relativistic constituent quark model within the diquark picture~\cite{Lu:2017meb},
and the chiral perturbation theory~\cite{Li:2017pxa}.

To end this work, we note that the electromagnetic transition of $\Xi^{*++}_{cc} \rightarrow \Xi^{++}_{cc}\gamma$ is probably the main decay mode of the $\Xi^{*++}_{cc}$, and its decay width is probably large enough for the $\Xi^{*++}_{cc}$ to be observed in the $\Xi^{++}_{cc}\gamma$ channel. Considering that the $\Xi^{++}_{cc}$ has been recently discovered by the LHC Collaboration~\cite{Aaij:2017ueg}, we propose to continually search for the doubly charmed baryon $\Xi^{*++}_{cc}$ of $J^P = 3/2^+$ in future LHCb and BelleII experiments.

\section*{ACKNOWLEDGMENTS}

This project is supported by
the National Natural Science Foundation of China under Grants No. 11722540, No. 11175073, No. 11205011, No. 11222547, No. 11375024, No. 11475015, No. 11575008, No. 11261130311, and No. 11621131001,
the 973 program,
the Ministry of Education of China (SRFDP under Grant No. 20120211110002 and the Fundamental Research Funds for the Central Universities),
the Fok Ying-Tong Education Foundation (Grant No. 131006), and
the National Program for Support of Top-notch Young Professionals.

\appendix

\section{Light-cone sum rule results}
\label{sec:result}

In this appendix we list the sum rules for the electromagnetic transition $\Xi^{*++}_{cc} \rightarrow \Xi^{++}_{cc}\gamma$, obtained by calculating $\Pi_\alpha(p, \, k , \, q , \, \epsilon)$ defined in Eq.~(\ref{eq:correlation}) at the quark and gluon level using the method of operator product expansion (OPE). The results are
\begin{widetext}
\begin{eqnarray}
\label{eq:sumrule1}
&& \Pi^{\Xi^{*++}_{cc} \rightarrow \Xi^{++}_{cc} \gamma}_\alpha(p, \, k , \, q , \, \epsilon)
= \Pi^{\Xi^{++}_{cc} \gamma \rightarrow \Xi^{*++}_{cc}}_{\alpha,light}(p, \, k , \, q , \, \epsilon) + \Pi^{\Xi^{++}_{cc} \gamma \rightarrow \Xi^{*++}_{cc}}_{\alpha,heavy}(p, \, k , \, q , \, \epsilon) \, ,
\\ && \Pi^{\Xi^{++}_{cc} \gamma \rightarrow \Xi^{*++}_{cc}}_{\alpha,light}(p, \, k , \, q , \, \epsilon)
\\ \nonumber &=&
\int d^4x \int {d^4p_1\over(2\pi)^4} \int {d^4p_2\over(2\pi)^4} \int_0^1 du \times e^{-i k x} e^{- i (1-u) qx} \times \epsilon^{\alpha\nu\rho\sigma}q_\nu \epsilon_\rho \times e_u \times \Big \{
\\ \nonumber && \cos\theta_1 t_1 \times \Big(
-\frac{   f_{3\gamma}  x_\sigma (3 m_c^2 + 2 p_1\cdot p_2)}{2 (p_1^2-m_c^2) (p_2^2-m_c^2)} \psi^{(a)}(u)
+\frac{   f_{3\gamma} x_\sigma p_1 \cdot p_2}{72 (p_1^2-m_c^2)^2 (p_2^2-m_c^2)^2} \langle g_s^2 G G \rangle \psi^{(a)}(u)
\\ \nonumber &&
-\frac{   f_{3\gamma} x_\sigma (3 p_1^2 + 2 p_1\cdot p_2)}{24 (p_1^2-m_c^2)^4 (p_2^2-m_c^2)} m_c^2 \langle g_s^2 G G \rangle \psi^{(a)}(u)
-\frac{   f_{3\gamma} x_\sigma (3 p_2^2 + 2 p_1\cdot p_2)}{24 (p_1^2-m_c^2) (p_2^2-m_c^2)^4} m_c^2 \langle g_s^2 G G \rangle \psi^{(a)}(u)
\Big)
\\ \nonumber && +\cos\theta_1 t_2 \times \Big(
-\frac{10i  \chi (p_{1\sigma} + p_{2\sigma})}{(p_1^2-m_c^2) (p_2^2-m_c^2)} m_c \langle \bar q q \rangle \phi_\gamma(u)
+\frac{5 i  \chi (p_{1\sigma} + p_{2\sigma})}{72 (p_1^2-m_c^2)^2 (p_2^2-m_c^2)^2} m_c \langle \bar q q \rangle \langle g_s^2 G G \rangle \phi_\gamma(u)
\\ \nonumber &&
-\frac{5 i  \chi (p_1^2 p_{2\sigma} + m_c^2 p_{1\sigma})}{6 (p_1^2-m_c^2)^4 (p_2^2-m_c^2)} m_c \langle \bar q q \rangle \langle g_s^2 G G \rangle \phi_\gamma(u)
-\frac{5 i  \chi (p_2^2 p_{1\sigma} + m_c^2 p_{2\sigma})}{6 (p_1^2-m_c^2) (p_2^2-m_c^2)^4} m_c \langle \bar q q \rangle \langle g_s^2 G G \rangle \phi_\gamma(u)
\Big)
\\ \nonumber && +\sin\theta_1 t_1 \times \Big(
-\frac{   f_{3\gamma} x_\sigma (3 m_c^2 + 2 p_1 \cdot p_2)}{4 (p_1^2-m_c^2) (p_2^2-m_c^2)} \psi^{(a)}(u)
-\frac{3 i  \chi (p_{1\sigma} + p_{2\sigma})}{(p_1^2-m_c^2) (p_2^2-m_c^2)} m_c \langle \bar q q \rangle \phi_\gamma(u)
\\ \nonumber &&
+\frac{   f_{3\gamma} x_\sigma p_1 \cdot p_2}{144 (p_1^2-m_c^2)^2 (p_2^2-m_c^2)^2} \langle g_s^2 G G \rangle \psi^{(a)}(u)
+\frac{  i  \chi (p_{1\sigma} + p_{2\sigma})}{48 (p_1^2-m_c^2)^2 (p_2^2-m_c^2)^2} m_c \langle \bar q q \rangle \langle g_s^2 G G \rangle \phi_\gamma(u)
\\ \nonumber &&
-\frac{   f_{3\gamma} x_\sigma (3 p_1^2 + 2 p_1 \cdot p_2)}{48 (p_1^2-m_c^2)^4 (p_2^2-m_c^2)} m_c^2 \langle g_s^2 G G \rangle \psi^{(a)}(u)
-\frac{  i  \chi (p_1^2 p_{2\sigma} + m_c^2 p_{1\sigma})}{4 (p_1^2-m_c^2)^4 (p_2^2-m_c^2)} m_c \langle \bar q q \rangle \langle g_s^2 G G \rangle \phi_\gamma(u)
\\ \nonumber &&
-\frac{   f_{3\gamma} x_\sigma (3 p_2^2 + 2 p_1 \cdot p_2)}{48 (p_1^2-m_c^2) (p_2^2-m_c^2)^4} m_c^2 \langle g_s^2 G G \rangle \psi^{(a)}(u)
-\frac{  i  \chi (p_2^2 p_{1\sigma} + m_c^2 p_{2\sigma})}{4 (p_1^2-m_c^2) (p_2^2-m_c^2)^4} m_c \langle \bar q q \rangle \langle g_s^2 G G \rangle \phi_\gamma(u)
\Big)
\\ \nonumber && +\sin\theta_1 t_2 \times \Big(
-\frac{3  f_{3\gamma} x_\sigma}{(p_1^2-m_c^2) (p_2^2-m_c^2)} m_c^2 \psi^{(a)}(u)
-\frac{10 i  \chi p_{2\sigma}}{(p_1^2-m_c^2) (p_2^2-m_c^2)} m_c \langle \bar q q \rangle \phi_\gamma(u)
\\ \nonumber &&
+\frac{ f_{3\gamma} x_\sigma}{24 (p_1^2-m_c^2)^2 (p_2^2-m_c^2)^2} m_c^2 \langle g_s^2 G G \rangle \psi^{(a)}(u)
+\frac{5 i  \chi p_{2\sigma}}{72 (p_1^2-m_c^2)^2 (p_2^2-m_c^2)^2} m_c \langle \bar q q \rangle \langle g_s^2 G G \rangle \phi_\gamma(u)
\\ \nonumber &&
-\frac{ f_{3\gamma} p_1^2 x_\sigma}{4 (p_1^2-m_c^2)^4 (p_2^2-m_c^2)} m_c^2 \langle g_s^2 G G \rangle \psi^{(a)}(u)
-\frac{5 i  \chi p_1^2 p_{2\sigma}}{6 (p_1^2-m_c^2)^4 (p_2^2-m_c^2)} m_c \langle \bar q q \rangle \langle g_s^2 G G \rangle \phi_\gamma(u)
\\ \nonumber &&
-\frac{ f_{3\gamma} p_2^2 x_\sigma}{4 (p_1^2-m_c^2) (p_2^2-m_c^2)^4} m_c^2 \langle g_s^2 G G \rangle \psi^{(a)}(u)
-\frac{5 i  \chi p_{2\sigma}}{6 (p_1^2-m_c^2) (p_2^2-m_c^2)^4} m_c^3 \langle \bar q q \rangle \langle g_s^2 G G \rangle \phi_\gamma(u)
\Big) \Big \}
\\ \nonumber &+&
\int d^4x \int {d^4p_1\over(2\pi)^4} \int {d^4p_2\over(2\pi)^4} \int_0^1 dw \int \mathcal{D}\underline{\alpha} \times e^{-i k x} e^{-i (\alpha_2 + w \alpha_3) qx} \times \epsilon^{\alpha\nu\rho\sigma}q_\nu \epsilon_\rho \times e_u \times m_c \langle \bar q q \rangle \times \Big \{
\\ \nonumber && \cos\theta_1 t_2 \times \Big(
-\frac{  i (3 p_{1\sigma} + p_{2\sigma})}{  (p_1^2-m_c^2)^2 (p_2^2-m_c^2)} S(\underline{\alpha})
+\frac{  i p_{2\sigma}}{  (p_1^2-m_c^2)^2 (p_2^2-m_c^2)} \widetilde S(\underline{\alpha})
-\frac{  i (5 p_{1\sigma} + 3 p_{2\sigma})}{  (p_1^2-m_c^2)^2 (p_2^2-m_c^2)} T_1(\underline{\alpha})
\\ \nonumber &&
+\frac{  i (5 p_{1\sigma} + 3 p_{2\sigma})}{  (p_1^2-m_c^2)^2 (p_2^2-m_c^2)} T_2(\underline{\alpha})
-\frac{3 i  p_{2\sigma}}{  (p_1^2-m_c^2)^2 (p_2^2-m_c^2)} T_3(\underline{\alpha})
+\frac{3 i  p_{2\sigma}}{  (p_1^2-m_c^2)^2 (p_2^2-m_c^2)} T_4(\underline{\alpha})
\\ \nonumber &&
-\frac{  i (p_{1\sigma} + 3 p_{2\sigma})}{  (p_1^2-m_c^2) (p_2^2-m_c^2)^2} S(\underline{\alpha})
+\frac{  i p_{1\sigma}}{  (p_1^2-m_c^2) (p_2^2-m_c^2)^2} \widetilde S(\underline{\alpha})
-\frac{  i (3 p_{1\sigma} + 5 p_{2\sigma})}{  (p_1^2-m_c^2) (p_2^2-m_c^2)^2} T_1(\underline{\alpha})
\\ \nonumber &&
+\frac{  i (3 p_{1\sigma} + 5 p_{2\sigma})}{  (p_1^2-m_c^2) (p_2^2-m_c^2)^2} T_2(\underline{\alpha})
-\frac{3 i  p_{1\sigma}}{  (p_1^2-m_c^2) (p_2^2-m_c^2)^2} T_3(\underline{\alpha})
+\frac{3 i  p_{1\sigma}}{  (p_1^2-m_c^2) (p_2^2-m_c^2)^2} T_4(\underline{\alpha})
\Big)
\\ \nonumber && +\sin\theta_1 t_1 \times \Big(
-\frac{  i (3 p_{1\sigma} + p_{2\sigma})}{2 (p_1^2-m_c^2)^2 (p_2^2-m_c^2)} S(\underline{\alpha})
-\frac{  i  p_{2\sigma}}{2 (p_1^2-m_c^2)^2 (p_2^2-m_c^2)} \widetilde S(\underline{\alpha})
-\frac{  i (3 p_{1\sigma} + p_{2\sigma})}{2 (p_1^2-m_c^2)^2 (p_2^2-m_c^2)} T_1(\underline{\alpha})
\\ \nonumber &&
+\frac{  i (3 p_{1\sigma} + p_{2\sigma})}{2 (p_1^2-m_c^2)^2 (p_2^2-m_c^2)} T_2(\underline{\alpha})
+\frac{  i  p_{2\sigma}}{2 (p_1^2-m_c^2)^2 (p_2^2-m_c^2)} T_3(\underline{\alpha})
-\frac{  i  p_{2\sigma}}{2 (p_1^2-m_c^2)^2 (p_2^2-m_c^2)} T_4(\underline{\alpha})
\\ \nonumber &&
-\frac{  i  (p_{1\sigma} + 3 p_{2\sigma})}{2 (p_1^2-m_c^2) (p_2^2-m_c^2)^2} S(\underline{\alpha})
-\frac{  i   p_{1\sigma}}{2 (p_1^2-m_c^2) (p_2^2-m_c^2)^2} \widetilde S(\underline{\alpha})
-\frac{  i  (p_{1\sigma} +  p_{2\sigma})}{2 (p_1^2-m_c^2) (p_2^2-m_c^2)^2} T_1(\underline{\alpha})
\\ \nonumber &&
+\frac{  i  (p_{1\sigma} +  p_{2\sigma})}{2 (p_1^2-m_c^2) (p_2^2-m_c^2)^2} T_2(\underline{\alpha})
+\frac{  i   p_{1\sigma}}{2 (p_1^2-m_c^2) (p_2^2-m_c^2)^2} T_3(\underline{\alpha})
-\frac{  i   p_{1\sigma}}{2 (p_1^2-m_c^2) (p_2^2-m_c^2)^2} T_4(\underline{\alpha})
\Big)
\\ \nonumber && +\sin\theta_1 t_2 \times \Big(
-\frac{  i  p_{2\sigma}}{  (p_1^2-m_c^2)^2 (p_2^2-m_c^2)} S(\underline{\alpha})
+\frac{  i  p_{2\sigma}}{  (p_1^2-m_c^2)^2 (p_2^2-m_c^2)} \widetilde S(\underline{\alpha})
-\frac{3 i  p_{2\sigma}}{  (p_1^2-m_c^2)^2 (p_2^2-m_c^2)} T_1(\underline{\alpha})
\\ \nonumber &&
+\frac{3 i  p_{2\sigma}}{  (p_1^2-m_c^2)^2 (p_2^2-m_c^2)} T_2(\underline{\alpha})
-\frac{3 i  p_{2\sigma}}{  (p_1^2-m_c^2)^2 (p_2^2-m_c^2)} T_3(\underline{\alpha})
+\frac{3 i  p_{2\sigma}}{  (p_1^2-m_c^2)^2 (p_2^2-m_c^2)} T_4(\underline{\alpha})
\\ \nonumber &&
-\frac{3 i  p_{2\sigma}}{  (p_1^2-m_c^2) (p_2^2-m_c^2)^2} S(\underline{\alpha})
-\frac{5 i  p_{2\sigma}}{  (p_1^2-m_c^2) (p_2^2-m_c^2)^2} T_1(\underline{\alpha})
+\frac{5 i  p_{2\sigma}}{  (p_1^2-m_c^2) (p_2^2-m_c^2)^2} T_2(\underline{\alpha})
\Big)\Big \} \, ,
\\ && \Pi^{\Xi^{++}_{cc} \gamma \rightarrow \Xi^{*++}_{cc}}_{\alpha,heavy}(p, \, k , \, q , \, \epsilon)
\\ \nonumber &=&
\int d^4x \int {d^4p_1\over(2\pi)^4} \int {d^4p_2\over(2\pi)^4} \times e^{-i k x} \times \epsilon^{\alpha\nu\rho\sigma}q_\nu \epsilon_\rho \times e_c \times \Big \{
\\ \nonumber &&  \cos\theta_1 t_1 \times \Big(
 \frac{12 x_{\sigma} (m_c^2 + p_1 \cdot p_2)}{(p_1^2-m_c^2)^2 (p_2^2-m_c^2) \pi^2 x^4}
+\frac{12 x_{\sigma} (m_c^2 + p_1 \cdot p_2)}{(p_1^2-m_c^2) (p_2^2-m_c^2)^2 \pi^2 x^4}
\Big)
+\cos\theta_1 t_2 \times \Big(
 \frac{2 i (3 p_{1\sigma} + p_{2\sigma})}{  (p_1^2-m_c^2)^2 (p_2^2-m_c^2)} m_c \langle \bar q q \rangle
 \\ \nonumber &&
-\frac{  i x^2 (3 p_{1\sigma} + p_{2\sigma})}{8 (p_1^2-m_c^2)^2 (p_2^2-m_c^2)} m_c \langle g_s \bar q \sigma G q \rangle
+\frac{2 i (p_{1\sigma} + 3 p_{2\sigma})}{  (p_1^2-m_c^2) (p_2^2-m_c^2)^2} m_c \langle \bar q q \rangle
-\frac{  i x^2 (p_{1\sigma} + 3 p_{2\sigma})}{8 (p_1^2-m_c^2) (p_2^2-m_c^2)^2} m_c \langle g_s \bar q \sigma G q \rangle
\Big)
\\ \nonumber && +\sin\theta_1 t_1 \times \Big(
-\frac{6 x_{\sigma} (m_c^2 - 2 p_1 \cdot p_2)}{(p_1^2-m_c^2)^2 (p_2^2-m_c^2) \pi^2 x^4}
+\frac{  i (3 p_{1\sigma} + p_{2\sigma})}{  (p_1^2-m_c^2)^2 (p_2^2-m_c^2)} m_c \langle \bar q q \rangle
-\frac{  i x^2 (3 p_{1\sigma} + p_{2\sigma})}{16 (p_1^2-m_c^2)^2 (p_2^2-m_c^2)} m_c \langle g_s \bar q \sigma G q \rangle
\\ \nonumber &&
+\frac{18 x_{\sigma} m_c^2}{(p_1^2-m_c^2) (p_2^2-m_c^2)^2 \pi^2 x^4}
+\frac{  i (p_{1\sigma} + 3 p_{2\sigma})}{  (p_1^2-m_c^2) (p_2^2-m_c^2)^2} m_c \langle \bar q q \rangle
-\frac{  i x^2 (p_{1\sigma} + 3 p_{2\sigma})}{16 (p_1^2-m_c^2) (p_2^2-m_c^2)^2} m_c \langle g_s \bar q \sigma G q \rangle
\Big)
\\ \nonumber && +\sin\theta_1 t_2 \times \Big(
\frac{48 x_{\sigma} m_c^2}{(p_1^2-m_c^2)^2 (p_2^2-m_c^2) \pi^2 x^4}
+\frac{2  i p_{2\sigma}}{  (p_1^2-m_c^2)^2 (p_2^2-m_c^2)} m_c \langle \bar q q \rangle
-\frac{  i x^2 p_{2\sigma}}{8 (p_1^2-m_c^2)^2 (p_2^2-m_c^2)} m_c \langle g_s \bar q \sigma G q \rangle
\\ \nonumber &&
+\frac{48 x_{\sigma} m_c^2}{(p_1^2-m_c^2) (p_2^2-m_c^2)^2 \pi^2 x^4}
+\frac{6  i p_{2\sigma}}{  (p_1^2-m_c^2) (p_2^2-m_c^2)^2} m_c \langle \bar q q \rangle
-\frac{3  i x^2 p_{2\sigma}}{8 (p_1^2-m_c^2) (p_2^2-m_c^2)^2} m_c \langle g_s \bar q \sigma G q \rangle
\Big)\Big \} \,  .
\end{eqnarray}
\end{widetext}
where $e_u$ and $e_c$ are the electric charges of the up and charm quarks, respectively. The integration limits are
\begin{equation}
\int \mathcal{D}\underline{\alpha} = \int_0^1 d\alpha_1 \int_0^1 d\alpha_2 \int_0^1 d\alpha_3 \delta(1-\alpha_1-\alpha_2-\alpha_3) \, .
\end{equation}
It can be naturally separated into two parts: $\Pi^{\Xi^{++}_{cc} \gamma \rightarrow \Xi^{*++}_{cc}}_{\alpha,light}(p, \, k , \, q , \, \epsilon)$ is obtained when the photon is omitted from the light quark, while $\Pi^{\Xi^{++}_{cc} \gamma \rightarrow \Xi^{*++}_{cc}}_{\alpha,heavy}(p, \, k , \, q , \, \epsilon)$ is obtained when the photon is omitted from the two charm quarks.

After performing the Borel transformation at both the hadron and quark-gluon levels, we obtain the following sum rules
\begin{eqnarray}
\label{eq:sumrule2}
\Pi_\alpha(s_0 , M_B)
&\approx& {2e\over3}~g~\epsilon^{\alpha\nu\rho\sigma}p_\nu q_\rho \epsilon_\sigma \times {f_{\Xi^{*++}_{cc}} f_{\Xi^{++}_{cc}}}
\\ \nonumber && \times \left( - e^{-\widetilde M^2 / M_B^2} + {2 \widetilde M^2 \over M_B^2 } e^{-\widetilde M^2 / M_B^2} \right)
\\ \nonumber &=& \int^{s_0}_{s_<} e^{-s/M_B^2} \rho(s) ds \, ,
\end{eqnarray}
where $s_0$ and $M_B$ are the threshold value and Borel mass, respectively. The spectral density $\rho(s)$ is
\begin{widetext}
\begin{eqnarray}
\rho(s) &=& \int_0^1 du \times \epsilon^{\alpha\nu\rho\sigma} p_\nu q_\rho \epsilon_\sigma \times \frac{e_u}{16 \pi^2} \times \int^{\alpha_{max}}_{\alpha_{min}}d\alpha \times \Big \{
3 ( 2 \cos\theta_1 t_1 +\sin\theta_1 t_1 ) \times f_{3\gamma}\psi^{(a)}(u) \times (1-\alpha) \alpha
\\ \nonumber &&
- 2 ( 10 \cos\theta_1 t_2 + 3 \sin\theta_1 t_1 + 5 \sin\theta_1 t_2 ) \times \chi \phi_\gamma(u) \times m_c\langle \bar q q \rangle \times (1-\alpha)
\Big \}
\\ \nonumber &&
+ \int_0^1 du \times \epsilon^{\alpha\nu\rho\sigma} p_\nu q_\rho \epsilon_\sigma \times \frac{e_u}{1152 \pi^2} \times \int^1_0d\alpha~\delta( s - \frac{m_c^2}{(1-\alpha)\alpha}) \times \Big \{
\\ \nonumber &&
(2\cos\theta_1 t_1 +\sin\theta_1 t_1 ) \times f_{3\gamma} \psi^{(a)}(u) \times
\Big( 180 m_c^2 -\langle g_s^2 G G \rangle \times \Big(\frac{1}{M_B^2} +\frac{(1-10 \alpha) m_c^2}{(1-\alpha)^2 \alpha M_B^4} +\frac{5 m_c^4}{(1-\alpha)^3 M_B^6} \Big) \Big)
\\ \nonumber &&
+6 \sin\theta_1 t_2 \times f_{3\gamma} \psi^{(a)}(u) \times
\Big( 72 m_c^2  -\langle g_s^2 G G \rangle \times \Big (\frac{(1-7\alpha) m_c^2}{(1-\alpha)^2 \alpha M_B^4}
+\frac{2 m_c^4}{(1-\alpha)^3 M_B^6} \Big)
\\ \nonumber &&
+( 10 \cos\theta_1 t_2 + 3 \sin\theta_1 t_1 + 5\sin\theta_1 t_2 ) \times \chi\phi_\gamma(u) \times  m_c \langle \bar q q \rangle \langle g_s^2 G G \rangle \times \Big(
\frac{7 \alpha-6}{\alpha^2 M_B^2}+\frac{2 (1-3\alpha+3\alpha^2) m_c^2}{(1-\alpha)^2 \alpha^3 M_B^4} \Big)
\Big \}
\\ \nonumber &+&
\int_0^1 dw \int \mathcal{D}\underline{\alpha} \times \epsilon^{\alpha\nu\rho\sigma} p_\nu q_\rho \epsilon_\sigma \times \frac{e_u}{16 \pi^2} \times \int^1_0d\alpha~\delta(s-\frac{m_c^2}{(1-\alpha) \alpha})  \times m_c \langle \bar q q \rangle \times \Big \{
(2 \cos\theta_1 t_2 + \sin\theta_1 t_2 )
\\ \nonumber && \times \Big(
\Big[3 S(\underline{\alpha}) + 5 T_1(\underline{\alpha}) - 5 T_2(\underline{\alpha}) \Big]
+\frac{1-\alpha}{\alpha} \Big[S(\underline{\alpha})-\tilde S(\underline{\alpha})+3 T_1(\underline{\alpha})
-3 T_2(\underline{\alpha})+3 T_3(\underline{\alpha})-3 T_4(\underline{\alpha})\Big]\Big )
\\ \nonumber && +\sin\theta_1 t_1 \times \Big(
3 \Big[S(\underline{\alpha})+T_1(\underline{\alpha})-T_2(\underline{\alpha})\Big]
+\frac{1-\alpha}{\alpha} \Big[S(\underline{\alpha})+\tilde S(\underline{\alpha})+T_1(\underline{\alpha})
-T_2(\underline{\alpha})-T_3(\underline{\alpha})+T_4(\underline{\alpha})\Big]\Big )
\\ \nonumber &+&
\epsilon^{\alpha\nu\rho\sigma} p_\nu q_\rho \epsilon_\sigma \times \frac{3 e_c}{32 \pi^4}  \times
\int^{\alpha_{max}}_{\alpha_{min}}\frac{d\alpha}{\alpha} \int^{\beta_{max}}_{\beta_{min}}d\beta \times \Big \{
(2 \cos\theta_1 t_1 +\sin\theta_1 t_1 )
\times (
m_c^2 - 2 m_c^2(\alpha+\beta) + \alpha\beta s )
\\ \nonumber &&
+8 \sin\theta_1 t_2 \times m_c^2 \times (1-\alpha-\beta)
\Big \}
\\ \nonumber &-&
\epsilon^{\alpha\nu\rho\sigma} p_\nu q_\rho \epsilon_\sigma \times \frac{e_c}{32 \pi^2} \times \int^1_0d\alpha~\delta(s-\frac{m_c^2}{(1-\alpha) \alpha}) \times
\Big( 2 \cos\theta_1 t_2 + \sin\theta_1 t_1 + \sin\theta_1 t_2 \Big)
\\ \nonumber && \times  m_c \times \Big(4 (2+\frac{1}{\alpha}) \langle \bar q q \rangle
+  (\frac{2}{M_B^2}+\frac{1}{\alpha M_B^2}+\frac{3 m_c^2}{\alpha M_B^4}+\frac{3 m_c^2}{(1-\alpha) M_B^4}+\frac{m_c^2}{\alpha^2 M_B^4}) \langle g_s \bar q \sigma G q \rangle \Big)
\, ,
\end{eqnarray}
\end{widetext}
where the integration limits are
\begin{eqnarray*}
\alpha_{min} &=& \frac{1-\sqrt{1-4m_c^2/s}}{2} \, ,
\\ \alpha_{max} &=& \frac{1+\sqrt{1-4m_c^2/s}}{2} \, ,
\\ \beta_{min} &=& \frac{\alpha m_c^2}{\alpha s-m_c^2} \, ,
\\ \beta_{max} &=& 1-\alpha \, .
\end{eqnarray*}

\section{Light-cone photon distribution amplitudes}
\label{sec:wavefunction}

There are many photon distribution amplitudes used in the present study. We list them in Eqs.~(\ref{eq:da1}-\ref{eq:da8}): $\phi_{\gamma}$ is the leading twist-2 distribution amplitude; $\psi^{(v)}$, $\psi^{(a)}$, ${\cal A}$, and ${\cal V}$ are the twist-3 ones; $h_{\gamma}$, $S$, $\widetilde{S}$, and $T_{1,2,3,4}$ are the twist-4 ones. We refer to Ref.~\cite{Ball:2002ps} for the detailed expressions of these photon distribution amplitudes, and to Refs.~\cite{ball} for more distribution amplitudes of pseudoscalar and vector mesons.

\begin{widetext}
\begin{eqnarray}
\langle 0 |\bar q (z) \gamma_{\mu} q (-z)| \gamma^{(\lambda)}(q)\rangle
\label{eq:da1}
&=& e_q\,f_{3\gamma}\, e^{(\lambda)}_{\perp\mu} \int_{0}^{1} \!du\,e^{i\xi qz}\, \psi^{(v)}(u, \mu)\, ,
\\ \langle 0|\bar q(z) \gamma_{\mu} \gamma_{5} q(-z)| \gamma^{(\lambda)}(q)  \rangle
&=& \frac12e_q\,f_{3\gamma}\, \varepsilon_{\mu \nu q z}\,e^{(\lambda)}_{\perp\nu} \int_{0}^{1} \!du\, e^{i\xi qz}\,\psi^{(a)}(u, \mu)\,,
\\ \langle 0| \bar q(z) \sigma_{\alpha\beta} q(-z) |\gamma^{(\lambda)}(q)\rangle
&=& i  \,e_q\,\chi\, \langle\bar q q\rangle  \left( q_\beta e^{(\lambda)}_\alpha- q_\alpha e^{(\lambda)}_\beta \right)  \int_0^1 \!du\, e^{i\xi qz}\, \phi_{\gamma}(u,\mu)
\\ \nonumber && + \frac i2 e_q\,\frac{\langle\bar q q\rangle}{qz} \left( z_\beta e^{(\lambda)}_\alpha -z_\alpha e^{(\lambda)}_\beta \right) \int_0^1 \!du\, e^{i\xi qz}\, h_{\gamma}(u,\mu)\, ,
\\ \langle 0 |\bar q(z) g\widetilde G_{\mu\nu}(vz)\gamma_\alpha\gamma_5   q(-z)|\gamma^{(\lambda)}(q) \rangle
&=& e_q \,f_{3\gamma}\, q_\alpha [q_\nu e^{(\lambda)}_{\perp\mu}   - q_\mu e^{(\lambda)}_{\perp\nu}] \int \mathcal{D}\underline{\alpha} {\cal A}(\underline{\alpha}) e^{-iqz\alpha_v}\,,
\\ \langle 0 |\bar q(z) g G_{\mu\nu}(vz)i\gamma_\alpha q(-z)|\gamma^{(\lambda)}(q) \rangle
&=& e_q \,f_{3\gamma}\, q_\alpha[q_\nu e^{(\lambda)}_{\perp\mu}   - q_\mu e^{(\lambda)}_{\perp\nu}] \int \mathcal{D}\underline{\alpha} {\cal V}(\underline{\alpha}) e^{-iqz\alpha_v}\,,
\\ \langle 0 | \bar q(z)g{G}_{\mu\nu}(vz) q(-z) | \gamma^{(\lambda)}(q) \rangle
&=& ie_q \,\langle\bar q q\rangle [q_\nu e^{(\lambda)}_{\perp\mu}-q_\mu e^{(\lambda)}_{\perp\nu}] \int \mathcal{D}\underline{\alpha} S(\underline{\alpha}) e^{-iqz\alpha_v}\,,
\\ \langle 0| \bar q(z)g\tilde{G}_{\mu\nu}(vz)i\gamma_5 q(-z) |\gamma^{(\lambda)}(q) \rangle
&=& ie_{q}\,\langle\bar q q\rangle [q_\nu e^{(\lambda)}_{\perp\mu} -q_\mu e^{(\lambda)}_{\perp\nu}] \int \mathcal{D}\underline{\alpha} \widetilde{S}(\underline{\alpha}) e^{-iqz\alpha_v}\,,
\\ \langle 0 | \bar q(z)\sigma_{\alpha \beta}g{G}_{\mu\nu}(vz) q(-z) |\gamma^{(\lambda)}(q) \rangle
\label{eq:da8}
&=& e_{q}\,\langle\bar q q\rangle [ q_\alpha e^{(\lambda)}_{\perp\mu}g^\perp_{\beta\nu} - q_\beta e^{(\lambda)}_{\perp\mu}g^\perp_{\alpha\nu} - q_\alpha e^{(\lambda)}_{\perp\nu}g^\perp_{\beta\mu} + q_\beta e^{(\lambda)}_{\perp\nu}g^\perp_{\alpha\mu} ] T_1(v,qz)
\\ \nonumber &&+ e_{q}\,\langle\bar q q\rangle [ q_\mu e^{(\lambda)}_{\perp\alpha}g^\perp_{\beta\nu} -q_\mu e^{(\lambda)}_{\perp\beta}g^\perp_{\alpha\nu} -q_\nu e^{(\lambda)}_{\perp\alpha}g^\perp_{\beta\mu} +q_\nu e^{(\lambda)}_{\perp\beta}g^\perp_{\alpha\mu} ] T_2(v,qz)
\\ \nonumber &&+ \frac{e_q\,\langle\bar q q\rangle}{qz}
[ q_\alpha q_\mu e^{(\lambda)}_{\perp\beta}z_\nu
     -q_\beta q_\mu e^{(\lambda)}_{\perp\alpha}z_\nu
     -q_\alpha q_\nu e^{(\lambda)}_{\perp\beta}z_\mu
     +q_\beta q_\nu e^{(\lambda)}_{\perp\alpha}z_\mu ]
      T_3(v,qz)
\\ \nonumber &&+ \frac{e_q\,\langle\bar q q\rangle}{qz}
    [ q_\alpha q_\mu e^{(\lambda)}_{\perp\nu}z_\beta
     -q_\beta q_\mu e^{(\lambda)}_{\perp\nu}z_\alpha
     -q_\alpha q_\nu e^{(\lambda)}_{\perp\mu}z_\beta
     +q_\beta q_\nu e^{(\lambda)}_{\perp\mu}z_\alpha ]
      T_4(v,qz)\,.
\end{eqnarray}
\end{widetext}

\end{document}